\documentstyle[12pt,epsfig,cite]{article}

\newlength{\dinwidth}
\newlength{\dinmargin}
\def\docnum#1{\hbox to \hsize{\hskip123mm\hbox{#1}\hss}}
\def\date#1{\edef\@temp{#1}\ifx\@temp\@empty\def\@temp{\today}\fi
\hbox to \hsize{\hskip123mm\hbox{\@temp}\hss}}
\def\title#1{\vskip 0.8in plus 2in\begin{center}%
{\Large\bf#1\par}\vskip1.5em\end{center}\vskip 1in}
\def\@makefnmark{\hbox{$^{\@thefnmark)}$}}
\def\author#1{
\setcounter{footnote}{0}\def\@currentlabel{}%
\begingroup\def\thefootnote{\arabic{footnote}}
\def\@makefnmark{\hbox{$^{\@thefnmark)}$}}
\global\@topnum\z@ \large\begin{center}{\lineskip.5em
\begin{tabular}[t]{c}#1\end{tabular}\par}
\end{center}\par\vskip1.5em\@thanks\endgroup}

\def\abstract{\vskip0.8in plus 3in\begin{center}{\large\bf Abstract}\end{center}\quotation}

\newcommand{\QG}   {{\bf Q}}
\newcommand{\QGz}  {{\bf Q}^0}
\newcommand{\QGzi} {{\bf Q}_i^0}
\newcommand{\QGi}  {{\bf Q}_i}
\newcommand{\PG}   {{\bf \phi}}
\newcommand{\betb} {{\bf \beta}}
\newcommand{\ppb}  {${\rm{p\bar{p}}}\;$}
\newcommand{\ppbf} {${\bf{p\bar{p}}}\;$}
\newcommand{\ee}   {e$^+$e$^-$}
\newcommand{\cc}   {\rm{c{\bar c}}}
\newcommand{\bb}   {\rm{b{\bar b}}}
\newcommand{\eeqq} {${\rm{e^+e^- \rightarrow q \bar{q}}}\;$}
\newcommand{\qi}   {{\bf q}_i}
\newcommand{\qj}   {{\bf q}_j}

\newcommand{\qk}   {{\bf q}_k}
\newcommand{\zum}  {{\rm{\Sigma}}}
\newcommand{\dint} {{\rm{d}}}

\newcommand{\E}    {{\rm{e}}}
\newcommand{\I}    {{\rm{i}}}

\begin{document}

\begin{titlepage}

\flushright{DFF 263/12/1996}
\flushright{December 1996}
\title{Universality of thermal hadron production in 
 pp, \ppbf and \ee collisions}

\centerline{\large{F. Becattini}} 
\vspace*{0.5cm}
\centerline{INFN Sezione di Firenze} 
\centerline{Largo E. Fermi 2, I-50125 Firenze} 
\centerline{e-mail: becattini@fi.infn.it}
\centerline{FAX ++39/55/229330}

\begin{abstract}
 It is shown that the hadron production in high energy 
 pp, \ppb and \ee collisions, calculated by assuming that particles originate 
 in hadron gas fireballs at thermal and partial chemical equilibrium, agrees 
 very well with the data. The temperature of the hadron gas fireballs, 
 determined by fitting hadron abundances, does not seem to depend on the 
 centre of mass energy and kind of reaction, having a nearly constant value 
 of about 170 MeV. 
 This finding supports a universal hadronization mechanism in all kinds of 
 reactions consisting in a parton-hadron transition at critical values of 
 temperature and pressure.
\end{abstract}

\vspace*{2cm}

\center{{\it To be published in the proceedings 
    of the \\ XXXIII Eloisatron Workshop on \\ Universality features in 
    multihadron production and the leading effect \\ 
    October 19-25 1996, Erice (Italy)}}

\end{titlepage} 
  
\section{Introduction}
 
 The thermodynamical approach to hadron production in hadronic collisions was
 originally introduced by Hagedorn \cite{hag} almost thirty years ago. One of the 
 most important phenomenological indication of a thermal-like multihadron 
 production in those reactions at high energy was found in the universal slope 
 of transverse mass (i.e. $m_T=\sqrt{p^2_T+m^2}$) spectra \cite{hag2} which is 
 nowadays still being used extensively in heavy ions reactions, although it has 
 been realized that tranverse collective motions of the hadron gas may significantly
 distort the primitive thermal $m_T$-spectrum \cite{heinz}. A much better probe 
 of the existence of locally thermalized sources in hadronic collisions is the 
 overall production 
 rate of single hadron species which, being a Lorentz-invariant quantity, is 
 not affected by local collective motions of the hadron gas. The analysis of 
 hadron production rates with the thermodynamical {\it ansatz} implies that 
 inter-species chemical equilibrium is attained, which is a tighter requirement 
 with respect to a thermal-kinetical intra-species equilibrium assumed in the 
 analysis of $m_T$ spectra. However, since hadron abundances can provide a cleaner 
 indication of chemical and thermal equilibrium with respect to transverse 
 mass spectra, we have focused our attention on the former issue.\\
 The smallness of the collision systems studied here requires 
 appropriate theoretical tools: in order to properly compare 
 theoretical predicted multiplicities to experimental ones, the use of 
 statistical mechanics in its canonical form is mandatory, that means 
 exact quantum numbers conservation is required, unlike in the
 grand-canonical formalism \cite{hag3}. It will be shown indeed that 
 particle average particle multiplicities in small systems are heavily 
 affected by conservation laws well beyond what the use of chemical 
 potentials predicts.\\
 The thermodynamical approach to hadron production in \ee collisions 
 \cite{beca} is here extended to pp and \ppb collisions; some assumptions which
 were made for \ee, in particular the restriction to two-jet events, are
 released; all calculations are performed with a bigger symmetry group 
 (actually by also taking into account the conservation of electric charge). 
 With this generalized model, we also update our results 
 for \ee collisions.

\section{The model}

 In ref. \cite{beca,tesi} a thermodynamical model of hadron production in 
 \ee collisions was developed on the basis of the following 
 assumption: the hadronic jets observed in the final state of a \eeqq 
 event must be identified with hadron gas phases having a collective 
 motion. This identification is valid at the decoupling time, when 
 hadrons stop interacting after their formation and (possibly) a short 
 expansion ({\em freeze-out}). Throughout this paper we will refer to 
 such hadron gas phases with a collective motion as {\em fireballs}, 
 following refs. \cite{hag,hag2}. Since most events in a \eeqq 
 reaction are two-jet events, it was assumed that two fireballs are 
 formed and that their internal properties, namely quantum numbers, 
 are related to those of the corresponding primary quarks. In the 
 so-called {\em correlated jet scheme} correlations between the 
 quantum numbers of the two fireballs were allowed beyond the simple 
 correspondance between the fireball and the parent quark quantum 
 numbers. This scheme turned out to be in better agreement with the 
 data than a correlation-free scheme \cite{beca}. \\ 
 The more complicated structure of a hadronic collision does not allow 
 a straightforward extension of this model. If the assumption of 
 hadron gas fireballs is maintained, the possibility of an arbitrary number 
 of fireballs with an arbitrary configuration of quantum numbers 
 should be taken into account. To be specific, let us define
 a vector $\QG = (Q,N,S,C,B)$ with integer components equal to the electric 
 charge, baryon number, strangeness, charm and beauty respectively. 
 We assume that the final state of a pp or a \ppb interaction consists 
 of a set of $N$ fireballs, each with its own four-vector $\beta_i=u_i/T_i$,
 where $T_i$ is the temperature and $u_i= (\gamma_i,{\bf \betb}_i \gamma_i)$ is 
 the four-velocity \cite{tous}, quantum numbers $\QGzi$ and volume in 
 the rest frame $V_i$. The quantum vectors $\QGzi$ must fulfill the 
 overall conservation constraint $\sum_{i=1}^N \QGzi = \QGz $ where 
 $\QGz$ is the vector of the initial quantum numbers, that is $\QGz = 
 (2,2,0,0,0)$ in a pp collision and $\QGz = (0,0,0,0,0)$ in a \ppb 
 collision.\\ 
 The invariant partition function of a single fireball is, by definition:

 \begin{equation}
     Z_i(\QGzi) = \sum_{\rm{states}} \, \E^{- \beta_i \cdot P_i} 
     \delta_{\QGi,\QGzi} \; ,
 \end{equation}
 where $P_i$ is its total four-momentum. The factor $\delta_{\QGi,\QGzi}$ is the 
 usual Kronecker tensor, which forces the sum to be performed only over the fireball 
 states whose quantum numbers $\QGi$ are equal to the particular set $\QGzi$. 
 It is worth emphasizing that this partition function corresponds to the {\em canonical}
 ensemble of statistical mechanics since only the states fulfilling a fixed chemical 
 requirement, as expressed by the factor $\delta_{\QGi,\QGzi}$, are involved in the 
 sum (1).\\
 By using the integral representation of $\delta_{\QGi,\QGzi}$:

 \begin{equation}
    \delta_{\QGi,\QGzi} = \frac{1}{(2\pi)^5} \int_{0}^{2\pi} \!\!\!\! \int_{0}^{2\pi}
       \!\!\!\! \int_{0}^{2\pi} \!\!\!\! \int_{0}^{2\pi} \!\!\!\! \int_{0}^{2\pi}  
        \!\! \dint^5 \phi \,\, \E^{\,\I\,(\QGzi - \QGi) \cdot \phi} \; ,
 \end{equation}
 Eq.~(1) becomes:

 \begin{equation}
     Z_i(\QGzi) = \sum_{\rm{states}} \frac{1}{(2\pi)^5} \int_{0}^{2\pi} 
      \!\!\!\! \ldots \int_{0}^{2\pi} \!\!\! \dint^5 \phi \,\, \E^{- \beta_i \cdot P_i} 
      \E^{\,\I\, (\QGzi -\QGi) \cdot \phi} .
 \end{equation} 
 This equation could also have been derived from the general expression of partition 
 function of systems with internal symmetry \cite{zf1,zf2} by requiring a U(1)$^5$ 
 symmetry group, each U(1) corresponding to a conserved quantum number; that was the 
 procedure taken in ref. \cite{beca}.\\ 
 The sum over states in Eq.~3 can be worked out quite 
 straightforwardly for a hadron gas of $N_B$ boson species and $N_F$ 
 fermion species. A state is specified by a set of occupation numbers 
 $\{n_{j,k}\}$ for each phase space cell $k$ and for each particle 
 species $j$. Since $P_i=\sum_{j,k} p_{k} n_{j,k}$ and $\QGi= 
 \sum_{j,k} \qj n_{j,k}$, where $\qj = (Q_j,N_j,S_j,C_j,B_j)$ is the 
 quantum number vector associated to the $j^{th}$ particle species, 
 the partition function (3) reads, after summing over states:  

 \begin{equation}
  Z_i(\QGzi) = \frac{1}{(2\pi)^5} \int \, \dint^5 \phi \,\, 
      \E^{\,\I\, \QGzi \cdot \phi} \exp \, [ {\sum_{j=1}^{N_B} \sum_k \, 
      \log \, (1 - \E^{-\beta_i \cdot p_{k} -\I \qj \cdot \phi})^{-1}}
      + {\sum_{j=1}^{N_F} \sum_k \, \log \, (1+\E^{-\beta \cdot p_{k} 
      -\I \qj \cdot \phi})}] \; .
 \end{equation}
 The last expression of the partition function is manifestly Lorentz-invariant
 because the sum over phase space is a Lorentz-invariant 
 operation which can be performed in any frame. The most suitable one 
 is the fireball rest frame, where the four-vector $\beta_i$ reduces 
 to: 

 \begin{equation}
 \beta_i= (\frac{1}{T_i},0,0,0)
 \end{equation}
 $T_i$ being the temperature of the fireball. Moreover, the sum over phase space cells 
 in Eq.~(4) can be turned into an integration over momentum space going to the 
 continuum limit:

 \begin{equation}
 \sum_{k} \longrightarrow (2J_j+1) \, \frac{V}{(2\pi)^3} \int \dint^3 p  \; ,
 \end{equation}  
 where $V$ is the fireball volume and $J_j$ the spin of the $j^{th}$ 
 hadron. As in previous studies on \ee collisions \cite{beca} and 
 heavy ions collisions \cite{raf}, we supplement the ordinary 
 statistical mechanics formalism with a strangeness suppression factor 
 $\gamma_s$ accounting for a partial strangeness phase space 
 saturation; actually the Boltzmann factor $\E^{-\beta \cdot p_{k}}$ of 
 any hadron 
 species containing $s$ strange valence quarks or antiquarks is 
 multiplied by $\gamma_s^{s}$. With the transformation (6) and 
 choosing the fireball rest frame to perform the integration, the sum 
 over phase space in Eq.~(4) becomes: 
 
 \begin{eqnarray}
  && \sum_{k} \, \log \, (1 \pm \E^{-\beta_i \cdot p_{k} -\I \qj
   \cdot \phi})^{\pm 1} \longrightarrow \nonumber \\
  && \frac{(2J_j+1)\,V_i}{(2\pi)^3} \! \int \dint^3 p \,\, \log \, 
    (1 \pm \gamma_s^{s_j} \E^{-\sqrt{p^2+m_j^2}/T_i -\I \qj \cdot \phi})^{\pm 1}
    \equiv V_i \, F_j(T_i,\gamma_s) 
 \end{eqnarray}
 where the upper sign is for fermions, the lower for bosons and $V_i$ 
 is the fireball volume in its rest frame; the function 
 $F_j(T_i,\gamma_s)$ is a shorthand notation of the momentum integral in 
 Eq.~(7). Hence, the partition function (4) can be written: 

 \begin{equation}
    Z_i(\QGzi) = \frac{1}{(2\pi)^5} \int \, \dint^5 \phi \,\, 
    \E^{\,\I\, \QGzi \cdot \phi} \exp \, [V_i \sum_j F_j(T_i,\gamma_s)] \; .
 \end{equation}   
 The mean number $\langle n_j \rangle_i$ of the $j^{th}$ particle 
 species in the $i^{th}$ fireball can be derived from $Z(\QGzi)$ by 
 multiplying the Boltzmann factor $\exp\,(-\sqrt{p^2+m_j^2}/T)$, in the 
 function $F_j$ in Eq.~(8) by a fictitious fugacity $\lambda_j$ and 
 taking the derivative of $\log Z_i(\QGzi,\lambda_j)$ with respect to 
 $\lambda_j$ at $\lambda_j = 1$: 
    
 \begin{equation}
   \langle n_j \rangle_i \,\, = \frac {\partial}{\partial \lambda_j} 
   \log Z_i(\QGzi,\lambda_j) \Big|_{\lambda_j=1}   \; .  
 \end{equation}
 The partition function $Z_i(\QGi,\lambda_j)$ supplemented with the 
 $\lambda_j$ factor is still a Lorentz-invariant quantity and so is 
 the mean number $\langle n_j \rangle_i$. From a more physical point of 
 view, this means that the average multiplicity of any hadron does not 
 depend on fireball collective motion, unlike its mean number in a 
 particular momentum state.\\        
 The overall average multiplicity of the $j^{th}$ hadron, for a set of $N$ fireballs in 
 a certain quantum configuration $\{\QG_1^0,\ldots,\QG_N^0\}$ is the sum of all mean numbers 
 of that hadron in each fireball:

 \begin{equation}
   \langle n_j \rangle \,\, 
   = \sum_{i=1}^N \frac {\partial}{\partial \lambda_j} 
   \log Z_i(\QGzi,\lambda_j) \Big|_{\lambda_j=1} \nonumber \\ 
   = \frac {\partial}{\partial \lambda_j} \log \prod_{i=1}^N 
     Z_i(\QGzi,\lambda_j) \Big|_{\lambda_j=1} \; .  
 \end{equation} 
 In general, as the quantum number configurations may fluctuate, 
 hadron production should be further averaged over all possible 
 fireballs configurations ${\QG_1^0,\ldots,\QG_N^0}$ fulfilling the 
 constraint $\sum_{i=1}^N \QGzi = \QGz$. To this end, suitable 
 weights $w(\QG_1^0,\ldots,\QG_N^0)$, representing the probability of 
 configuration $\{\QG_1^0,\ldots,\QG_N^0\}$ to occur for a set of $N$ 
 fireballs, must be introduced. Basic features of those weights are:

 \begin{eqnarray}
     && w(\QG_1^0,\ldots,\QG_N^0) = 0  \qquad{\rm{if}}\quad  
          \sum_{i=1}^N \QGzi \neq \QGz \,, \nonumber \\
     && \sum_{\QG_1^0,\ldots,\QG_N^0} \!\!\!\! w(\QG_1^0,\ldots,\QG_N^0) = 1 \,.
 \end{eqnarray}
 For the overall average multiplicity of hadron $j$ we get:

 \begin{equation}
 \langle\!\langle n_j \rangle\!\rangle = \sum_{\QG_1^0,\ldots,\QG_N^0} 
   \!\!\!\! w(\QG_1^0,\ldots,\QG_N^0) \frac {\partial}{\partial \lambda_j} \log 
  \prod_{i=1}^N Z_i(\QGzi,\lambda_j) \Big|_{\lambda_j=1}  \; .
 \end{equation}  
 There are infinitely many possible choices of the weights 
 $w(\QG_1^0,\ldots,\QG_N^0)$, all of them equally legitimate. However, 
 one of them is the most pertinent from the statistical mechanics point 
 of view, namely: 
 
 \begin{equation}
     w(\QG_1^0,\ldots,\QG_N^0) = \frac{\delta_{\zum_i \QGzi,\QGz} \prod_{i=1}^N 
      Z_i(\QGzi)}{\sum_{\QG_1^0,\ldots,\QG_N^0} \!\!\! \delta_{\zum_i \QGzi,\QGz} 
      \prod_{i=1}^N Z_i(\QGzi)}  .    
 \end{equation}
 It can be shown indeed that this choice corresponds to the minimal deviation 
 from statistical equilibrium of the system as a whole. In fact, putting weights (13) 
 in the Eq.~(12), one obtains:

 \begin{equation}  
 \langle\!\langle n_j \rangle\!\rangle = \frac {\partial}{\partial \lambda_j} 
        \log \!\!\!\! \sum_{\QG_1^0,\ldots,\QG_N^0}  \!\!\!\! 
        \delta_{\zum_i \QGzi,\QGz} \prod_{i=1}^N 
        Z_i(\QGzi,\lambda_j)\Big|_{\lambda_j=1}. \!\!\!\!  
 \end{equation}   
 This means that the average multiplicity of any hadron can be derived 
 from the following function of $\QGz$: 

 \begin{equation}
 Z(\QGz) = \!\!\!\! \sum_{\QG_1^0,\ldots,\QG_N^0} \!\!\!\! \delta_{\zum_i \QGzi,\QGz} 
                    \prod_{i=1}^N Z_i(\QGzi) \; , 
 \end{equation} 
 with the same recipe given for a single fireball in Eq.~(9). By using 
 expression (1) for the partition functions $Z_i(\QGzi)$, Eq.~(15) 
 becomes: 

 \begin{equation}
 Z(\QGz)= \!\!\!\! \sum_{\QG_1^0,\ldots,\QG_N^0} \!\!\!\! \delta_{\zum_i \QGzi,\QGz} 
  \prod_{i=1}^N \sum_{\rm{states}_i} \, \E^{- \beta_i \cdot P_i} \delta_{\QGzi,\QGi} \; .                   
 \end{equation} 
 Since

 \begin{equation}
  \sum_{\QG_1^0,\ldots,\QG_N^0} \!\!\!\! \delta_{\zum_i \QGzi,\QGz} \,\, 
   \delta_{\QGzi,\QGi} = \delta_{\zum_i \QGi,\QGz} \; , 
 \end{equation} 
 the function (16) can be written as

 \begin{equation}
  Z(\QGz)=\!\!\!\! \sum_{\rm{states}_1} \!\! \ldots \!\! 
  \sum_{\rm{states}_N} \, \E^{- \beta_1 \cdot P_1} \ldots \E^{- 
  \beta_N \cdot P_N} \delta_{\zum_i \QGi,\QGz} \; .                    
 \end{equation}  
 This expression demonstrates that $Z(\QGz)$ may be properly called 
 the {\em global partition function} of a system split into $N$ 
 subsystems which are in mutual chemical equilibrium but not in mutual 
 thermal and mechanical equilibrium. Indeed it is a Lorentz-invariant 
 quantity and, in case of complete equilibrium, i.e. 
 $\beta_1=\beta_2=\ldots=\beta_N\equiv \beta$, it would reduce to:

 \begin{equation}
  Z(\QGz) =\sum_{\rm{states}_1} \!\! \ldots \!\! \sum_{\rm{states}_N}
  \, \E^{- \beta \cdot (P_1 + \ldots \cdot P_N)} \delta_{\zum_i \QGi,\QGz} \nonumber \\
  = \sum_{\rm{states}} \, \E^{- \beta \cdot P} \delta_{\QG,\QGz}  \; ,                  
 \end{equation}    
 which is the basic definition of the partition function.\\
 To summarize, the choice of weights (13) allows the construction of a 
 system which is out of equilibrium only by virtue of its subdivision 
 into several parts having different temperatures and velocities.
 Another very important consequence of that choice is the following: 
 if we assume that the freeze-out temperature of the various fireballs 
 is constant, that is $T_1 = \ldots = T_N \equiv T$, and that the 
 strangeness suppression factor $\gamma_s$ is constant too, then the 
 global partition function (18) has the following expression: 

 \begin{equation}
   Z(\QGz) = \frac{1}{(2\pi)^5} \int \, \dint^5 \phi \,\, 
    \E^{\,\I\, \QGz \cdot \phi} \exp \, [ (\zum_i V_i) \sum_j 
    F_j(T,\gamma_s)] \; .
 \end{equation}
 Here the $V_i$'s are the fireball volumes in their own rest frames; a 
 proof of (20) is given in ref. \cite{tesi}. Eq.~(20) demonstrates that 
 the global partition function has the same functional form (3), (4), 
 (8) as the partition function of a single fireball, once the volume $V_i$ 
 is replaced by the {\em global volume} $V \equiv \sum_{i=1}^N V_i$.
 Note that the global volume absorbs any dependence of the global 
 partition function (20) on the number of fireballs $N$. Thus, 
 possible variations of the number $N$ and the size $V_i$ of fireballs 
 on an event by event basis can be turned into fluctuations of the 
 global volume. In the remainder of this section these fluctuations 
 will be ignored; in Sect. 3 it will be shown that they do 
 not affect any of the following results on the average hadron
 multiplicities.\\    
 The average multiplicity of the $j^{th}$ hadron can be determined 
 with the formulae (14)-(15), by using expression (20) for the function 
 $Z(\QGz)$: 
 
 \begin{eqnarray}
   \langle\!\langle n_j \rangle\!\rangle & = & \frac{1}{(2\pi)^5}
    \int \dint^5 \phi \,\, \E^{\,\I\, \QGz \cdot \phi} \exp \, 
    [ V \sum_j F_j(T,\gamma_s)] \nonumber \\
   &\times & \frac {(2J_j+1)\,V}{(2\pi)^3} 
    \int \frac {\dint^3 p}{\gamma_s^{-s_j}
    \exp \,(\sqrt{p^2+m^2_j}/T+\I \qj \cdot \PG) \pm 1} \; ,
 \end{eqnarray}   
 where the upper sign is for fermions and the lower for bosons. This 
 formula can be written in a more compact form as a series: 

 \begin{equation}  
   \langle\!\langle n_j \rangle\!\rangle =  
    \sum_{n=1}^{\infty} (\mp 1)^{n+1} \,\gamma_s^{n s_j} 
    z_{j(n)} \, \frac{Z(\QGz-n\qj)}{Z(\QGz)} \; ,
 \end{equation}
 where the functions $z_{j(n)}$ are defined as:

 \begin{equation}  
    z_{j(n)} \equiv (2J_j+1)\, \frac{V}{(2\pi)^3} \int \dint^3 
   p \, \exp \, (-n \sqrt{p^2+m^2_j}/T) 
     = (2J_j+1) \, \frac{VT}{2\pi^2 n} \, m_j^2 \, 
      {\rm{K}}_2(\frac{nm_j}{T}) \; .  
 \end{equation}
 K$_2$ is the McDonald function of order 2. Eq.~(22) is the final 
 expression for the average multiplicity of hadrons at freeze-out. 
 Accordingly, the production rate of a hadron species depends only 
 on its spin, mass, quantum numbers and strange quark content.
 The {\em chemical factors} $Z(\QGz-n\qj)/Z(\QGz)$ in Eq.~(22) are a 
 typical feature of the canonical approach due to the requirement of exact 
 conservation of the initial set of quantum numbers. These factors 
 suppress or enhance production of particles according to the 
 vicinity of their quantum numbers to the initial $\QGz$ vector.
 The behaviour of $Z(\QG)$ as a function of electric charge, baryon number 
 and strangeness for suitable $T$, $V$ and $\gamma_s$ values is shown 
 in Fig. 1; for instance, it is evident that the baryon chemical 
 factors $Z(0,N,0,0,0)/Z(0,0,0,0,0)$ connected with an initially 
 neutral system play a major role in determining the baryon  
 multiplicities. The ultimate physical reason of ``charged" particle
 ($\qj \neq 0$) suppression with respect to ``neutral" ones ($\qj = 
 0$), in a completely neutral system ($\QGz = 0$), is the necessity, 
 once a ``charged" particle is created, of a simultaneous creation of 
 an anti-charged particle in order to fulfill the conservation laws. 
 In a {\em finite system} this pair creation mechanism is the more 
 unlikely the more massive is the lightest particle needed to 
 compensate the first particle's quantum numbers. For instance, once a 
 baryon is created, at least one anti-nucleon must be generated, which 
 is rather unlikely since its mass is much greater than the temperature 
 and the total energy is finite. On the other hand, if a non-strange 
 charged meson is generated, just a pion is needed to balance the 
 total electric charge; its creation is clearly a less unlikely event 
 with respect to the creation of a baryon as the energy to be spent is 
 lower. This argument illustrates why the dependence of $Z(\QG)$ on 
 the electric charge is much milder that on baryon number and 
 strangeness (see Fig. 1). In view of that, the dependence of 
 $Z(\QG)$ on electric charge was neglected in the previous study on 
 hadron production in \ee collisions \cite{beca}. 
 These chemical suppression effects are not accountable in a 
 grand-canonical framework; in fact, in a completely neutral system, 
 all chemical potentials should be set to zero and consequently 
 ``charged" particles do not undergo any suppression with respect to 
 ``neutral" ones.

\section{Other constraints and approximations}

 Actually, the global partition function (18) has to be modified in \ppb 
 collisions owing to a major effect in such reactions, the {\it leading 
 baryon effect} \cite{zichi}. Indeed, the sum (18) includes states 
 with vanishing net absolute value of baryon number, whereas in \ppb 
 collisions at least one baryon-antibaryon pair is always observed. 
 Hence, the simplest way to account for the leading baryon effect is to 
 exclude those states from the sum. Thus, if $|N| = \sum_i |N_i|$ 
 denotes the absolute value of the baryon number of the system, the 
 global partition function (18) should be turned into: 

 \begin{equation}
   Z = \!\!\!\! \sum_{\rm{states}_1} \!\! \ldots \!\! 
  \sum_{\rm{states}_N} \, \E^{- \beta_1 \cdot P_1} \ldots \E^{- 
  \beta_N \cdot P_N} \delta_{\zum_i \QGi,\QGz} 
   - \sum_{\rm{states}_1} \!\! \ldots \!\! 
  \sum_{\rm{states}_N} \, \E^{- \beta_1 \cdot P_1} \ldots \E^{- 
  \beta_N \cdot P_N} \delta_{\zum_i \QGi,\QGz} \delta_{|N|, 0}   \; .           
 \end{equation}
 The first term, that we define as $Z_1(\QGz)$, is equal to the 
 function $Z(\QGz)$ in Eqs.~(18), (20), while the second term is the 
 sum over all states having vanishing net absolute value of baryon 
 number. The absolute value of baryon number can be treated as a new 
 independent quantum number so that the processing of the partition 
 function described in Eqs.~(1)-(3) can be repeated for the second 
 term in Eq.~(24) with a U(1)$^6$ symmetry group. Accordingly, this 
 term can be naturally denoted by $Z_2(\QGz,0)$, so that Eq.~ (24) 
 reads: 
 
 \begin{equation}
  Z = Z_1(\QGz) - Z_2(\QGz,0)  \; .
 \end{equation}
 By using the integral representation of $\delta_{|N|, 0}$ 

 \begin{equation}
    \delta_{|N|, 0} = \frac{1}{2\pi} \int_{0}^{2\pi} \!\! \dint \psi 
      \,\, \E^{\,\I\, |N| \cdot \psi}
 \end{equation}
 in the second term of Eq.~(24), one gets:

 \begin{eqnarray}
    Z_2(\QGz,0) & = & \frac{1}{(2\pi)^6} \int  \dint^5 \phi \,\, 
     \E^{\,\I\, \QGz \cdot \phi} \exp [V \sum_j F_j(T,\gamma_s)] \nonumber \\
     &\times & \int \dint \psi \,\, \exp [\sum_j 
     \frac{(2J_j+1)V}{(2\pi)^3} \int \dint^3 p \, \log \, (1 + \gamma_s^{s_j} 
     \E^{-\sqrt{p^2+m_j^2}/T -\I \qj \cdot \phi - \,\I\, \psi})] \;,  
 \end{eqnarray}
 where the first sum over $j$ runs over all mesons and the second over 
 all baryons. The average multiplicity of any hadron species can be 
 derived from the global partition function (25) with the usual 
 prescription: 

 \begin{equation}
   \langle\!\langle n_j \rangle\!\rangle = \frac {\partial}{\partial \lambda_j} 
         \log Z(\lambda_j) \Big|_{\lambda_j=1}  \; .
 \end{equation}
 The calculation of the average multiplicity of primaries according to 
 Eq.~(22) (and Eq.~(28)) involves several rather complicated five-dimensional 
 integrals which have been calculated numerically after some useful 
 approximations, described in the following. Since the temperature is 
 expected to be below 200 MeV, the primary production rate of all 
 hadrons, except pions, is very well approximated by the first term of 
 the series (22): 

 \begin{equation}  
   \langle\!\langle n_j \rangle\!\rangle \simeq  
   \gamma_s^{s_j}\, z_j \, \frac{Z(\QGz-\qj)}{Z(\QGz)} \; ,
 \end{equation}  
 where we have put $z_j \equiv z_{j(1)}$. This approximation corresponds 
 to the Boltzmann limit of Fermi and Bose statistics. Actually, for a 
 temperature of 170 MeV, the primary production rate of K$^+$, the 
 lightest hadron after pions, differs at most (i.e. without the
 strangeness suppression parameter and the chemical factors which further 
 reduce the contribution of neglected terms) by 1.5\% from that 
 calculated with Eq.~(29), well within usual experimental 
 uncertainties. Corresponding Boltzmannian approximations can be made 
 in the function $Z(\QG)$, namely

 \begin{equation}
  \log \, (1\pm \E^{-\sqrt{p^2+m^2_j}/T -\I 
    \qj \cdot \phi })^{\pm 1} \simeq \E^{-\sqrt{p^2+m^2_j}/T -\I 
    \qj \cdot \phi } \; ,
 \end{equation}
 which turns Eq.~(20) (for a generic $\QG$) into:

 \begin{equation}
      Z(\QG) \simeq \frac{1}{(2\pi)^5} \int \, \dint^5 \phi \,\, 
       \E^{\,\I\, \QG \cdot \phi} \exp \, [ \sum_j z_j \gamma_s^{s_j} 
       \E^{-\I \qj \cdot \phi} 
          + \sum_{j=1}^{3} \frac{V}{(2\pi)^3} \int \dint^3 p \, 
      \log \, (1 - \E^{-\sqrt{p^2+m_j^2}/T -\I \qj \cdot \phi})^{-1}] \; ,
 \end{equation} 
 where the first sum runs over all hadrons except pions and the second 
 over pions.\\
 As a further consequence of the expected temperature value, the $z$ 
 functions of all charmed and bottomed hadrons are very small: with 
 $T= 170$ MeV and a primary production rate of K mesons of the order 
 of one, as the data states, the $z$ function of the lightest charmed 
 hadron, D$^0$, turns out to be $\approx 10^{-4}$; chemical factors 
 produce a further suppression of a factor $\approx 10^{-4}$. 
 Therefore, thermal production of heavy flavoured hadrons can be 
 neglected, as well as their $z$ functions in the exponentiated sum in 
 Eq.~(31), so that the integration over the variables $\phi_4$ and 
 $\phi_5$ can be performed: 

 \begin{eqnarray}
   Z(\QG,C,B) &\simeq& \frac{1}{(2\pi)^3} \int \, \dint^3 \phi 
   \,\, \E^{\,\I\, \QG \cdot \phi} \exp \, [\sum_j z_j \gamma_s^{s_j} 
   \E^{-\I \qj \cdot \phi} \nonumber \\
    &-& \sum_{j=1}^{3} \frac{V}{(2\pi)^3} 
   \int \dint^3 p \, \log \,(1 - \E^{-\sqrt{p^2+m_j^2}/T -\I \qj \cdot \phi})] \, 
       \delta_{C,0} \delta_{B,0} \equiv 
      \zeta(\QGz) \, \delta_{C,0} \delta_{B,0} \; . 
 \end{eqnarray} 
 $\QG$ and $\qj$ are now three-dimensional vectors consisting of
 electric charge, baryon number, and strangeness; the five-dimensional 
 integrals have been reduced to three-dimensional ones.\\ 
 Apart from the hadronization contribution, which is expected to be 
 negligible in this model, production of heavy flavoured hadrons in 
 high energy collisions mainly proceeds from hard perturbative QCD 
 processes of c${\rm{\bar{c}}}$ and b${\rm{\bar{b}}}$ pairs 
 creation. The fact that promptly generated heavy quarks do not 
 reannihilate into light quarks indicates a strong deviation from 
 statistical equilibrium of charm and beauty, much stronger than   
 the strangeness suppression linked with $\gamma_s$. Nevertheless, it 
 has been found in \ee collisions \cite{beca} that the relative 
 abundances of charmed and bottomed hadrons are in agreement with 
 those predicted by the statistical equilibrium assumption, confirming 
 its full validity for light quarks and quantum numbers associated to 
 them. The additional source of heavy flavoured hadrons arising from 
 perturbative processes can be accounted for by modifying the partition 
 function (31). In particular, the presence of one heavy flavoured 
 hadron and one anti-flavoured hadron should be demanded in a fraction 
 of events $R_c = \sigma (I \rightarrow \cc) / \sigma 
 (I)$ (or $R_b = \sigma (I \rightarrow \bb) / \sigma (I)$) where 
 $I$ is the initial colliding system, namely \ee, pp or \ppb, and 
 $\sigma(I)$ is meant to be the hadronic, inelastic and
 non-single-diffractive cross section respectively. Accordingly, the 
 partition function to be used in events with a perturbative c${\rm{\bar{c}}}$ 
 pair, is, by analogy with Eq.~(24)-(25) and the leading baryon effect: 

 \begin{eqnarray}
   Z & = & \!\!\!\! \sum_{\rm{states}_1} \!\! \ldots \!\! \sum_{\rm{states}_N}
  \, \E^{- \beta_1 \cdot P_1} \ldots \E^{- \beta_N \cdot P_N} \delta_{\zum_i \QGi,\QGz} \nonumber \\
  & - & \sum_{\rm{states}_1} \!\! \ldots \!\! \sum_{\rm{states}_N}
  \, \E^{- \beta_1 \cdot P_1} \ldots \E^{- \beta_N \cdot P_N} \delta_{\zum_i \QGi,\QGz}
  \delta_{|C|, 0} \equiv Z_1(\QGz) - Z_2(\QGz,0) \: ,            
 \end{eqnarray} 
 where $|C|$ is the absolute value of charm. The partition function to be
 used in events with a perturbative b${\rm{\bar{b}}}$ is, {\it mutatis mutandis},
 analogous. 
 Therefore, in the following analysis, we limit ourselfs to the 
 case of a c${\rm{\bar{c}}}$ pair, being understood that all derived results hold 
 for the b${\rm{\bar{b}}}$ case as well.\\
 In order to derive the average multiplicity of charmed hadrons in events in 
 which one $\cc$ pair is created owing to a hard QCD process, the 
 usual Eq.~(28) should be used with the partition function $Z$ in Eq.~(33). The 
 $Z$ can be worked out the same way as for the leading baryon effect, namely:

 \begin{equation}
 Z = Z_1(\QGz)-Z_2(\QGz,0) \; .
 \end{equation}
 where the function $Z_1$ can be written as in Eq.~(31):

 \begin{equation}
   Z_1(\QG) \simeq \frac{1}{(2\pi)^5} \int \, \dint^5 \phi \,\, \E^{\,\I\, \QG \cdot \phi} 
       \exp \, [ \sum_j z_j \gamma_s^{s_j} \E^{-\I \qj \cdot \phi} 
   + \sum_{j=1}^{3} \frac{V}{(2\pi)^3} \int \dint^3 p \, 
      \log \, (1 - \E^{-\sqrt{p^2+m_j^2}/T -\I \qj \cdot \phi})^{-1}] \; ,
 \end{equation} 
 and the function $Z_2$:

 \begin{eqnarray}
   Z_2(\QGz,K) & = & \frac{1}{(2\pi)^6} \int  \dint^5 \phi \,\, \int \dint \psi 
      \,\, \E^{\,\I\, K \psi} \exp \, [ \sum_{j=1} z_j \gamma_s^{s_j} 
      \E^{-\I \qj \cdot \phi -\I |C_j| \psi}  \nonumber \\ 
    & + & \sum_{j=1}^{3} \frac{V}{(2\pi)^3} \int \dint^3 p \, 
      \log \, (1 - \E^{-\sqrt{p^2+m_j^2}/T -\I \qj \cdot \phi})^{-1}] \; ,
 \end{eqnarray}
 where the second sum in the exponentials in both Eqs.~(35) and (36) runs over the 
 three pion states and $|C_j|$ in Eq.~(36) is the absolute value of $j^{th}$ hadron's 
 charm.\\ 
 Henceforth $\QGz$ and $\qi$, $\qj$, $\qk$ are to be understood as three-dimensional 
 vectors having as components electric charge, baryon number and strangeness, while 
 charm and beauty will be explicitely written down. By using this notation, the 
 average multiplicity of a charmed hadron with $C_j = 1$ turns out to be (see Eq.~(29)):

 \begin{equation} 
 \langle\!\langle n_j \rangle\!\rangle 
          =  z_j \frac{Z_1(\QGz-\qj,-1,0) - Z_2(\QGz-\qj,-1,0,-1)}
                      {Z_1(\QGz,0,0) - Z_2(\QGz,0,0,0)} \; .  
 \end{equation}
 Since the $z$ functions of heavy flavoured hadrons are $\ll 1$ a power expansion 
 in the $z_j$'s of all charmed and anticharmed hadrons can be performed from 
 $z_j = 0$ in the integrands of Eqs.~(35) and (36), that is: 

 \begin{equation}
  \exp [\sum_j \gamma_s^{s_j} z_j \E^{-\I \qj \cdot \phi}] \simeq 
   1 + \sum_j \gamma_s^{s_j} z_j \E^{-\I \qj \cdot \phi} 
   + \frac{1}{2} \sum_{i,j} \gamma_s^{s_i} \gamma_s^{s_j} z_i z_j 
   \E^{-\I (\qj+\qi) \cdot \phi}   
 \end{equation}
 for Eq.~(35) and

 \begin{equation}
   \exp \, [\sum_j \gamma_s^{s_j} z_j \E^{-\I \qj \cdot \phi -\I |C_j| \psi}] 
  \simeq  1 + \sum_j z_j \E^{-\I \qj \cdot \phi - \,\I\, |C_j| \psi} 
   + \frac{1}{2} \sum_{i,j} \gamma_s^{s_i} \gamma_s^{s_j} z_i z_j 
   \E^{-\I (\qj+\qi) \cdot \phi -2 \I \, |C_j| \psi} 
 \end{equation}
 for Eq.~(36). Furthermore, the $z$ functions of the bottomed hadrons can be 
 negelected as they are $\ll 1$ as well and beauty in Eq.~(37) is always set to zero.\\
 Those expansions permit carrying out integrations in the variables $\psi$, 
 $\phi_4$ and $\phi_5$ in Eqs.~(35) and (36). Thus:

 \begin{equation}
   Z_1(\QGz-\qj,-1,0) \simeq \sum_i \gamma_s^{s_i} z_i \zeta(\QGz-\qj-\qi)  \; , 
 \end{equation}
 where the sum runs over the {\em anticharmed hadrons} as the integration in $\phi_4$ 
 of terms associated to charmed hadrons yields zero. The $\zeta$ function on the 
 right-hand side is the same as in Eq.~(32). Moreover: 

 \begin{equation}
   Z_1(\QGz,0,0) \simeq \zeta(\QGz) + \sum_{i,k} \gamma_s^{s_i} \gamma_s^{s_k}
   z_i z_k \zeta(\QGz-\qi-\qk) \; ,
 \end{equation}
 where the index $i$ runs over all charmed hadrons and index $k$ over all anti-charmed
 hadrons.\\
 Owing to the presence of the absolute value of charm in the exponential 
 $\exp[\,\I\, |C_j| \psi]$ in its integrand function, the function $Z_2(\QGz,C,B,K)$ 
 vanishes if $K \le 0$ and yields $K^{th}$-order terms of the power expansion in 
 $z_j$ if $K \ge 0$ (see Eq.~(36)). Therefore: 
  
 \begin{equation}
   Z_2(\QGz,0,0,0) = \zeta(\QGz)  
 \end{equation}
 and   

 \begin{equation}
   Z_2(\QGz,-1,0,-1) = 0  \; .
 \end{equation}
 Finally, inserting Eqs.~(40), (41), (42) and (43) in Eq.~(37) one gets:  

 \begin{equation}
   \langle\!\langle n_j \rangle\!\rangle 
   = \gamma_s^{s_j} z_j \,\, \frac{\sum_i \gamma_s^{s_i }z_i \zeta(\QGz-\qj-\qi)}
    {\sum_{i,k} \gamma_s^{s_i} \gamma_s^{s_k} z_i z_k \zeta(\QGz-\qi-\qk)} \; ,
 \end{equation}
 where the indices $j$, $k$ label charmed hadrons and $i$ labels anticharmed hadrons. 
 From previous equation it results that the overall number of charmed hadrons 
 is 1, as it must be if c quark production from fragmentation is negligible.
 The average multiplicity of anticharmed hadrons is of course equal to charmed 
 hadrons one. The same formula (44) holds for the average multiplicity of 
 bottomed hadrons in events with a perturbatively generated $\bb$ pair.
 If leading baryon effect is taken into account for \ppb collisions, the formula 
 (44) gets more complicated, but the procedure is essentially the same.\\
 It is clear that a possible charm or beauty suppression parameter $\gamma_c$ 
 or $\gamma_b$, introduced by analogy with strangeness suppression parameter 
 $\gamma_s$, would not be revealed from the study of heavy flavoured hadron production
 because a single factor multiplying all $z_j$ functions would cancel from the 
 ratio in the right-hand side of Eq.~(44). \\
 So far, we tacitly assumed that the parameters $T$, $V$ and $\gamma_s$ 
 do not fluctuate on an event by event basis. If freeze-out occurs at a 
 fixed hadronic density in all events, then it is a reasonable {\it 
 ansatz} that $T$ and $\gamma_s$ do not undergo any fluctuation since 
 the density mainly depends on those two variables. However, there 
 could still be significant volume fluctuations due to event by event 
 variations of the number and size of the fireballs from which the 
 primary hadrons emerge (see end of Sect. 2). However, it can be shown 
 that, as far as the average hadron multiplicity 
 ratios are concerned, possible fluctuations of $V$ are negligible. To 
 this end, let us define $\rho(V)$ as the probability density of 
 picking a volume between $V$ and $V+\dint V$ in a single event. The 
 primary average multiplicity of the $j^{th}$ hadron is then:
   
 \begin{equation}  
    \langle\!\langle n_j \rangle\!\rangle
      = \!\! \int \dint V \rho (V) \sum_{n=1}^{\infty} (\mp 1)^{n+1} 
      \, \gamma_s^{n s_j} z_{j(n)} \, \frac{Z(\QGz-n\qj)}{Z(\QGz)} \; .
 \end{equation}     
 Provided that the volume $V$ does not fluctuate over a too large range, 
 the chemical factors can be taken out of the integral in Eq.~(45) and 
 evaluated at the mean volume $\overline V$, because their dependence 
 on it is mild and can be neglected (see Fig. 2). Therefore, the 
 integrand depends on the volume only through the functions $z_{j(n)}$ 
 whose dependence on $V$ is linear (see Eq.~(23)) and which can be 
 reexpressed as

 \begin{equation}
 z_{j(n)}(V,T,\gamma_s) = V \, \xi_{j(n)}(T,\gamma_s)     \; .
 \end{equation}
 Then, from Eq.~(45),

 \begin{equation}  
   \langle\!\langle n_j \rangle\!\rangle \simeq \sum_{n=1}^{\infty} 
      (\mp 1)^{n+1} \, \gamma_s^{n s_j} \xi_{j(n)}(T,\gamma_s) \, 
      \frac{Z(\QGz-n\qj)}{Z(\QGz)} \int \dint V \rho (V) \, V  \; .
 \end{equation}
 The integral on the right-hand side is the mean volume $\overline V$. Thus:

 \begin{equation}  
   \langle\!\langle n_j \rangle\!\rangle \simeq \sum_{n=1}^{\infty} 
   (\mp 1)^{n+1} \, \gamma_s^{n s_j} z_{j(n)}(\overline V,T,\gamma_s) 
   \, \frac{Z(\QGz-n\qj)}{Z(\QGz)} \; .  
 \end{equation} 
 It turns out that the relative hadron abundances do not depend on 
 the volume fluctuations since the mean volume $\overline V$ appearing in the 
 above equation (replacing the volume $V$ in Eq.~(23)) is the same for 
 all species: all results previously obtained are unaffected. 

\section{Fit procedure and results}

 The model described so far has three free parameters: the temperature 
 $T$, the global volume $V$ and the strangeness suppression parameter 
 $\gamma_s$. They have to be determined by a fit to the available data 
 on hadron inclusive production at each centre of mass energy. 
 Eq.~(28) yields the mean number of hadrons emerging directly from the 
 thermal source at freeze-out, the so-called primary hadrons, 
 \cite{beca,giova} as a function of the three free parameters. After 
 freeze-out, primary hadrons trigger a decay chain process which must 
 be properly taken into account in a comparison between model 
 predictions and experimental data, as the latter generally embodies 
 both primary hadrons and hadrons generated by heavier particles 
 decays. Therefore, in order to calculate overall average 
 multiplicities to be compared with experimental data, the primary 
 yield of each hadron species is added to the contribution stemming from 
 the decay of heavier hadrons, which is calculated by using 
 experimentally known decay modes and branching ratios \cite{pdg,jet}.
\\ The primary average multiplicity is calculated by using Eq.~(22)
 for pp collisions and Eq.~(28) for \ppb collisions. For \ee collisions
 the primary production rate of the $j^{th}$ hadron is given by:

 \begin{equation}
   \langle\!\langle n_j \rangle\!\rangle =
   \sum_{i=1}^5 R_i \frac{\partial}{\partial \lambda_j} 
   \log Z_i \Big|_{\lambda_j=1}   
 \end{equation}
 where:

 \begin{equation}
   R_i = \frac{\sigma({\rm e}^+{\rm e}^-\rightarrow{\rm q}_i\overline{\rm q}_i)}
         {\sigma({\rm e}^+{\rm e}^-\rightarrow{\rm hadrons})}
 \end{equation} 
 are the branching fractions into specific flavours and $Z_i$ are the 
 corresponding partition functions, namely Eq.~(32) for 
 u, d quarks, Eq.~(33) for c quarks and also for s and b quarks where 
 $|C|$ is replaced by $|B|$ and $|S|$ respectively. The fractions $R_i$ 
 depend on the centre of mass energy of the colliding system and determine 
 the primary production of heavy flavoured hadrons to be taken into account as 
 explained in the previous section.
 \\ As far as hadronic collisions are concerned,
 the fractions $R_c$ and $R_b$ are worst known with respect to \ee 
 collisions. Available data on charm cross-sections \cite{charm} indicate 
 a fraction $R_c \approx 10^{-2} \div 10^{-3}$ at centre of mass energies 
 $< 30$ GeV and, consequently, much lower values for bottom quark production. 
 Therefore, the perturbative production of heavy quarks in hadronic 
 collisions can be neglected as long as one deals with light flavoured 
 hadron production at $\sqrt s < 30$ GeV. We assume that it may be neglected 
 at any centre of mass energy. An upper estimate of charm production 
 in \ppb collisions at 900 GeV ($R_c=0.3$) based on theoretical calculations
 \cite{vogt}, does not affect significantly the fitted parameters.
 All light flavoured hadrons and resonances with a mass $< 1.7$ GeV 
 have been included among the primary generated hadron species. All
 heavy flavoured hadrons included in the JETSET 7.4 \cite{jet} program 
 tables have been used. 
 \\ The mass of resonances with $\Gamma > 1$ MeV has been 
 distributed according to a relativistic Breit-Wigner function within 
 $\pm 2\Gamma$ from the central value. The $\gamma_s$ strangeness 
 suppression factor has also been applied to neutral mesons such as 
 $\phi$, $\omega$, etc. according to the their strange valence quark
 content; mixing angles quoted in \cite{pdg} have been used. Once the
 average multiplicities of the primary hadrons have been calculated as 
 a function of the three parameters $T$, $V$ and $\gamma_s$, the decay 
 chain is performed until $\pi$, $\mu$, K$^{\pm}$, K$^0$, $\Lambda$, 
 $\Xi$, $\Sigma^{\pm}$, $\Omega^-$ or stable particles are reached, in 
 order to match the average multiplicity definition in pp and \ppb 
 collisions experiments. The decay chain is further extended in
 \ee collisions since experiments also include the decay products
 of K$^0_s$, $\Lambda$, $\Xi$, $\Sigma^{\pm}$ and $\Omega^-$ in their 
 multiplicity definition.
 Finally, the overall yield is compared with experimental measurements, 
 and the $\chi^2$: 

 \begin{equation} 
  \chi^2 = \sum_i ({\rm{theo}}_i-{\rm{expe}}_i)^2/{\rm{error}}_i^2 
 \end{equation}
 is minimized.\\
 As far as the data set is concerned, we used all available measurements
 of hadron multiplicities in non-single-diffractive 
 \ppb \cite{datappb} and inelastic pp collisions \cite{datapp} down 
 to a centre of mass energy of about 19 GeV and \ee collisions between
 29 and 91.2 GeV \cite{eedata}.\\
 Whenever several measurements at the same centre of mass energy have been 
 available, averages have been calculated according to a weighting 
 procedure described in ref. \cite{dean} prescribing rescaling of 
 errors to take into account {\em a posteriori} correlations and 
 disagreements of experimental results. If an experiment repeated
 the same measurement at the same centre of mass energy, we used
 only the most recent value.\\
 Since the decay chain is an essential step of the fitting procedure, 
 calculated theoretical multiplicities are affected by experimental 
 uncertainties on masses, widths and branching ratios of all involved 
 hadron species. In order to estimate the effect of these 
 uncertainties on the results of the fit, a two-step procedure for the fit 
 itself has been adopted: firstly, the fit has been performed with a 
 $\chi^2$  including only experimental errors and a set of parameters 
 $T_0$, $V_0$, $\gamma_{s0}$ has been obtained. Then, the various masses, 
 widths and branching ratios have been varied in turn by their errors, 
 as quoted in ref. \cite{pdg}, and new theoretical multiplicities 
 calculated, keeping the parameters $T_0$, $V_0$, $\gamma_{s0}$ fixed. 
 The differences between old and new theoretical multiplicity values 
 have been considered as additional systematic errors to be added in 
 quadrature to experimental errors. Finally, the fit has been 
 repeated with a $\chi^2$ including overall errors so as to obtain 
 final values for model parameters and for theoretical multiplicities. 
 Among the mass, width and branching ratio uncertainties, only 
 those producing significant variations of final hadron yields 
 (actually more than 130) have been considered.\\
 The fitted values of the parameters $T$, $V$, $\gamma_s$ at 
 various centre of mass energy points are quoted in Table 1.
 The fit quality is very good at almost all centre of mass 
 energies as demonstrated by the low values of $\chi^2$'s and by the 
 Figs. 3, 4, 5, 6, 7. Owing to the relatively large value of $\chi^2$ 
 at $\sqrt s = 27.4$ GeV in pp collisions and at $\sqrt s = 91.2$ GeV
 in \ee collisions, variations of fitted parameters larger than fit 
 errors must be expected when repeating the fit excluding data points 
 with the largest deviations from the theoretical values. Therefore, 
 the fit at $\sqrt s = 27.4$ GeV pp collisions has been repeated 
 excluding in turn ($\Delta^0$, $\rho^0$, $\phi$) and (K$^{-}$, 
 pions), respectively, from the data set; the fit at $\sqrt s = 91.2$ 
 GeV \ee collisions has been repeated excluding $\Sigma^{*\pm}$. The
 maximum difference between the new and old fit parameters has been 
 considered as an additional systematic error and is quoted in Table 
 1 within brackets. 
\begin{table}[t]
\caption{Values of fitted parameters in pp, \ppb and \ee collisions.}
\vspace{0.2cm}
\begin{center}
\begin{tabular}{|c c c c c|}
 $\sqrt s$ (GeV)            & $T$(MeV)              & $V T^3$                  &  $\gamma_s$                & $\chi^2/$dof \\ 
\noalign{\smallskip}\hline
\noalign{\smallskip}
   pp collisions          & & & &  \\
\noalign{\smallskip}\hline
\noalign{\smallskip} 
   19.5                     &$190.8\pm27.4$         & $5.8\pm3.1$              & $0.463\pm0.037$            &  6.38/4       \\
   23.8                     &$194.4\pm17.3$         & $6.3\pm2.5$              & $0.460\pm0.067$            &  2.43/2       \\  
   26.0                     &$159.0\pm9.5$          & $13.4\pm2.7$             & $0.570\pm0.030$            &  1.86/2       \\ 
   27.5                     &$169.0\pm2.1$          & $11.04\pm0.69$           & $0.510\pm0.011$            &  136.4/27     \\ 
                            &  $(\pm3.4)$           &   $(\pm1.4)$             &  $(\pm 0.025)$             &               \\
\noalign{\smallskip}\hline
\noalign{\smallskip}
   \ppb collisions          & & & &  \\         
\noalign{\smallskip}\hline
\noalign{\smallskip}
  200                       &$175.4\pm14.8$         & $24.3\pm7.9$             & $0.537\pm0.066$            &  0.698/2   \\ 
  546                       &$181.7\pm17.7$         & $28.5\pm10.4$            & $0.557\pm0.052$            &  3.80/1     \\ 
  900                       &$170.2\pm11.8$         & $43.2\pm11.8$            & $0.578\pm0.063$            &  1.79/2     \\ 
\noalign{\smallskip}\hline
\noalign{\smallskip}
   \ee collisions           & & & &  \\  
\noalign{\smallskip}\hline
\noalign{\smallskip} 
   $29\div 30$              &$163.6\pm3.6$          & $15.2\pm1.4$             & $0.724\pm0.045$            &  24.7/13     \\
   $34\div 35$              &$165.2\pm4.4$          & $14.6\pm1.6$             & $0.788\pm0.045$            &  10.5/8      \\  
   $42.6\div 44$            &$169.6\pm9.5$          & $14.7\pm3.1$             & $0.730\pm0.060$            &  4.94/4      \\
   91.2                     &$160.6\pm1.7$          & $26.4\pm1.3$             & $0.675\pm0.020$            &  60.8/21     \\ 
                            & $(\pm 3.1)$           & $(\pm 1.1)$              & $(\pm 0.027)$              &              \\
\noalign{\smallskip}\hline
\end{tabular}
\end{center}
\end{table}
 The fitted temperatures are compatible with a constant value at 
 freeze-out independently of collision energy and kind of 
 reaction (see Fig. 8) and it is in good agreement with that found in heavy ions 
 collisions \cite{satz}. The $\gamma_s$ parameter exhibits a very 
 slow rise from 20 to 900 GeV (see Fig. 9); its value of $\simeq 0.5$ over 
 the whole explored centre of mass energy range in hadronic collisions
 proves that complete strangeness equilibrium is not attained. 
 Furthermore, $\gamma_s$ appears to be definitely lower in pp and \ppb 
 collisions than in \ee collisions at equal centre of mass energy.\\
 On the other hand, the global volume does increase as a 
 function of centre of mass energy as it is proportional, for nearly 
 constant $T$ and $\gamma_s$, to overall multiplicity which indeed 
 increases with energy. Its values range from 6.4 fm$^3$ at $\sqrt s = 
 19.4$ GeV pp collisions, at a temperature of 191 MeV, up to 67 fm$^3$ 
 at $\sqrt s = 900$ GeV \ppb collisions at a temperature of 170 MeV.\\  
 Once $T$, $V$ and $\gamma_s$ are determined by fitting average 
 multiplicities of some hadron species, their values can be used to 
 predict average multiplicities of any other species, at a given 
 centre of mass energy.\\   
 The physical significance of the results found so far depends 
 on their stability as a function of the various approximations and 
 assumptions which have been introduced. First, the temperature and 
 $\gamma_s$ values are low enough to justify the use of the Boltzmann 
 limits (29), (30) for all hadrons except pions, as explained in Sect. 3. 
 As far as the effect of a cut-off in the hadronic mass spectrum 
 goes, the most relevant test proving that our results do not 
 depend on it is the stability of the number of primary hadrons 
 against changes of the cut-off mass. The fit procedure intrinsically 
 attempts to reproduce fixed experimental multiplicities; if the 
 number of primary hadrons does not change significantly by repeating 
 the fit with a slightly lower cut-off, the production of heavier 
 hadrons excluded by the cut-off must be negligible, in particular 
 with regard to its decay contributions to light hadron yields. In 
 this spirit all fits have been repeated moving the mass cut-off value 
 from 1.7 down to 1.3 GeV in steps of 0.1 GeV, checking the stability 
 of the amount of primary hadrons as well as of the fit parameters. 
 It is  worth remarking that the number of hadronic states with a mass 
 between 1.7 and 1.6 GeV is 238 out of 535 overall, so that their 
 exclusion is really a severe test for the reliability of the final 
 results. Figure 10 shows the model parameters and the primary hadrons 
 in \ppb collisions at $\sqrt s = 900$ GeV; above a cut-off of 1.5 GeV 
 the number of primary hadrons settles at an asymptotically stable 
 value, whilst the fitted values for $T$, $V$, $\gamma_s$ do not show 
 any particular dependence on the cut-off. Therefore, we conclude that 
 the chosen value of 1.7 GeV ensures that the obtained results are 
 meaningful. 

\section{Conclusions}

 A thermodynamical approach to hadronization, based on the use of the 
 canonical formalism of statistical mechanics ensuring the exact quantum 
 numbers conservation, succeeds in reproducing the average multiplicities
 of hadron species in \ee, pp and \ppb collisions. These quantities are best 
 suited in investigating the main characteristics of hadronization 
 because of their independence from local collective flows. The success
 of this statistical approach indicates that, apart from the violation 
 of strangeness phase space saturation as implemented in the $\gamma_s$ 
 parameter, hadron production in elementary high energy collisions is 
 dominated by phase space rather than by dynamical effects. The dynamics 
 is essentially contained in hadron gas fireballs collective motions 
 reflecting local hard partons kinematics.\\
 The temperature at freeze-out seems to be constant independently of kind 
 of reaction and centre of mass energy. This uniformity suggests that 
 parton-hadron transition cannot occur before the parameters of prehadronic
 matter such as density or pressure have dropped below critical values
 corresponding to a temperature of about 170 MeV in a (partially) 
 equilibrated hadron gas.
 This value is in agreement with lattice QCD results \cite{qm96}.\\
 From previous results it turns out that strangeness suppression, as well 
 as the survival of perturbatively created $\cc$ and $\bb$ pairs are the 
 only trace to strong interaction dynamics before hadronization. 
 This finding suggests that such non-equilibrium effects are mainly 
 related to quark mass thresholds.

\section*{Acknowledgments}
 I am greatly indebted with U. Heinz for his advice and his keen interest in this
 work. I wish to thank A. Giovannini, R. Hagedorn and H. Satz for 
 fruitful discussions. I warmly thank the organizers for their kind ospitality.

\newpage

\section*{Figure captions}
      
\begin{itemize}

\medskip 

\item[\rm Figure 1]
 Behaviour of the global partition function $Z$ as a 
 function of electric charge, baryon number and strangeness, keeping 
 all remaining quantum numbers set to zero, for $T = 170$ MeV, $V = 
 20$ fm$^3$ and $\gamma_s = 0.5$
                
\item[\rm Figure 2]
 Behaviour of the non-strange baryon chemical factor 
 $Z(0,1,0,0,0)/Z(0,0,0,0,0)$ as a function of volume for different 
 values of the temperature and a fixed value $\gamma_s =0.5$ for the 
 strangeness suppression parameter

\item[\rm Figure 3]
 Results of hadron multiplicity fits for pp collisions at
 centre of mass energies 19.4, 23.8 and 26 GeV. Experimental average 
 multiplicities are plotted versus calculated ones. The dashed lines 
 are the quadrant bisectors: well fitted points tend to lie on these lines

\item[\rm Figure 4]  
 Results of hadron multiplicity fit for pp collisions 
 at centre of mass energy of 27.4 GeV. Top: the experimental average 
 multiplicities 
 are plotted versus the calculated ones. The dashed line is the 
 quadrant bisector; well fitted points tend to lie on this line. 
 Bottom: residual distributions
 
\item[\rm Figure 5]  
 Results of hadron multiplicity fit for \ppb collisions at 
 centre of mass energies 200, 546 and 900 GeV. Experimental average 
 multiplicities are 
 plotted versus calculated ones. The dashed lines are the quadrant bisectors: 
 well fitted points tend to lie on these lines

\item[\rm Figure 6]  
 Results of hadron multiplicity fits for \ee collisions at 
 centre of mass energies 29, 35 and 44 GeV. Experimental average multiplicities are 
 plotted versus calculated ones. The dashed lines are the quadrant bisectors: 
 well fitted points tend to lie on these lines

\item[\rm Figure 7]  
 Results of hadron multiplicity fits for \ee collisions at 
 centre of mass energy of 91.2 GeV. Experimental average multiplicities are 
 plotted versus calculated ones. The dashed lines are the quadrant bisectors: 
 well fitted points tend to lie on these lines

\item[\rm Figure 8]  
 Freeze-out temperature values as a function 
 of centre of mass energy. The error bars within horizontal ticks for 
 27.4 GeV pp collisions and for 91.2 GeV \ee collisions are the fit 
 errors; the overall error bars are the sum in 
 quadrature of the fit error and the systematic error related to data 
 set variation (see text)   

\item[\rm Figure 9]  
 Strangeness suppression parameters $\gamma_s$ 
 as a function of centre of mass energy. The error bars within horizontal 
 ticks for 27.4 GeV pp collisions and for 91.2 GeV \ee collisions are the fit 
 errors; the overall error bars are the sum in quadrature of the fit 
 error and the systematic error related to data set variation (see text)   

\item[\rm Figure 10]  
 Dependence of fitted parameters and primary average 
 multiplicities (top) on the mass cut-off in the hadron mass spectrum
 for \ppb collisions at a centre of mass energy of 900 GeV 

\end{itemize}

\newpage
   
\begin{figure}[htbp]
\mbox{\epsfig{file=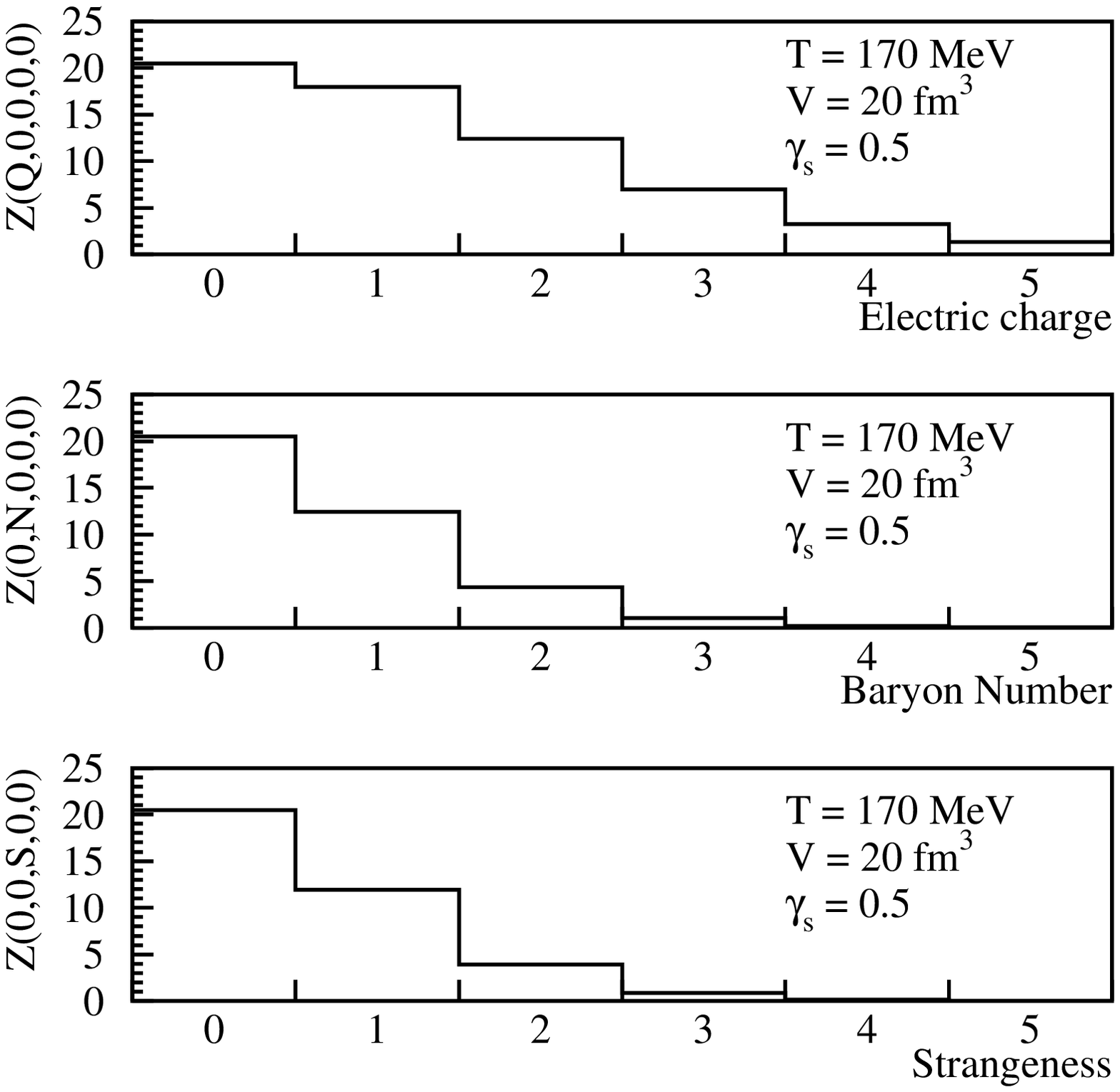,width=17cm}}
\caption{}
\end{figure}  

\newpage  
 
\begin{figure}[htbp]
\mbox{\epsfig{file=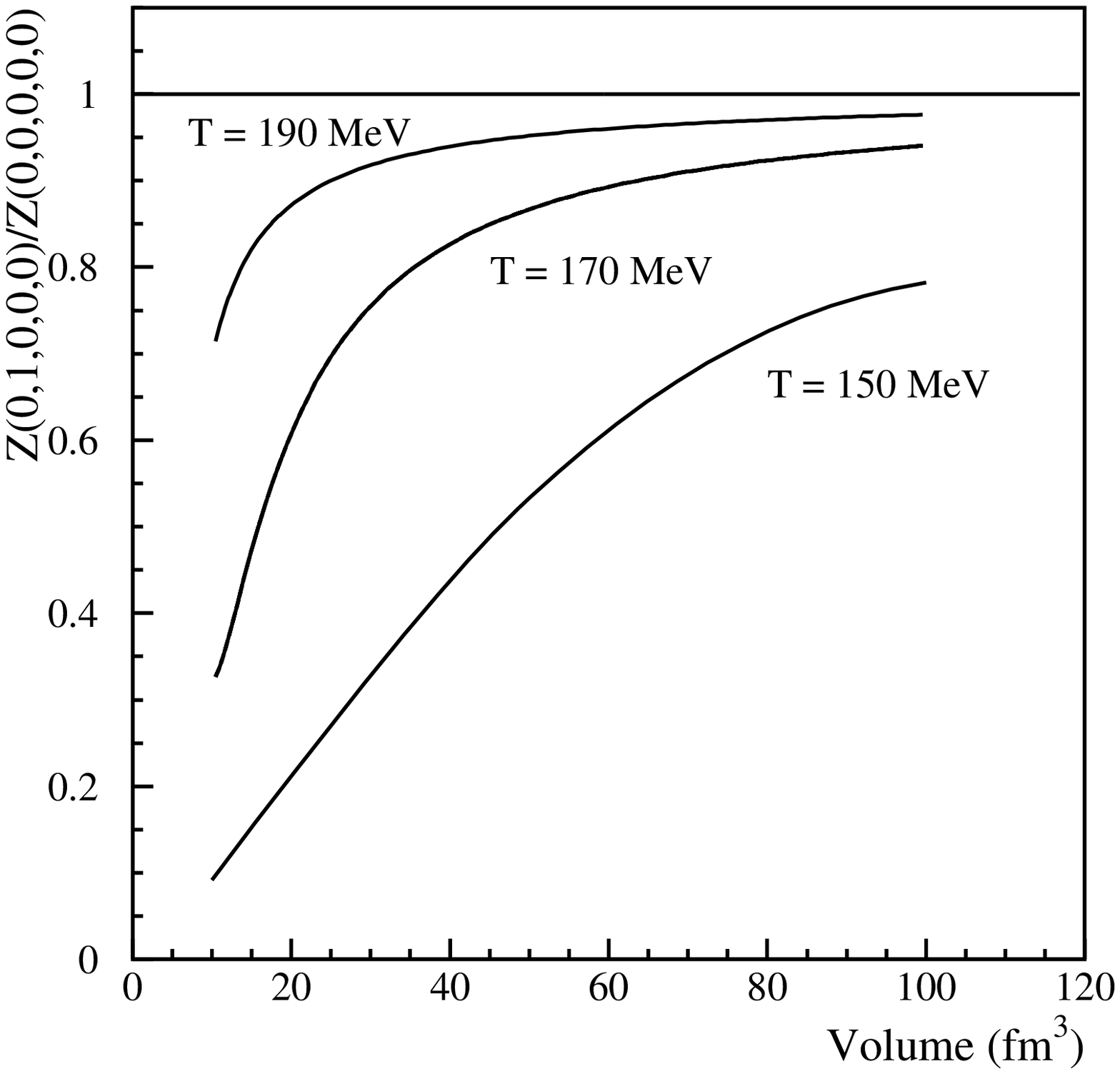,width=17cm}}
\caption{}
\end{figure}  

\newpage
              
\begin{figure}[htbp]
\mbox{\epsfig{file=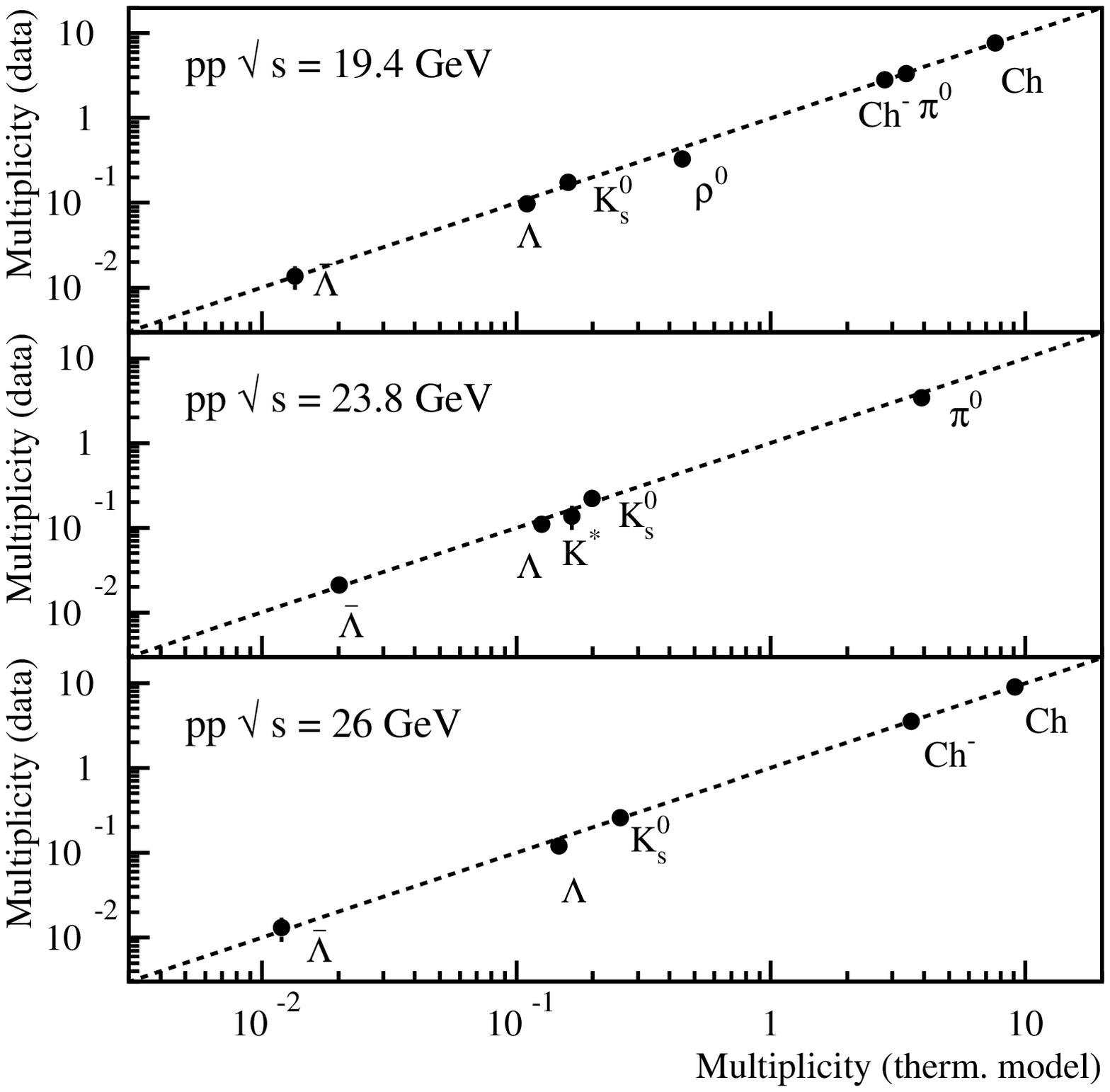,width=17cm}} 
\caption{}
\end{figure}

\newpage  
              
\begin{figure}[htbp]
\mbox{\epsfig{file=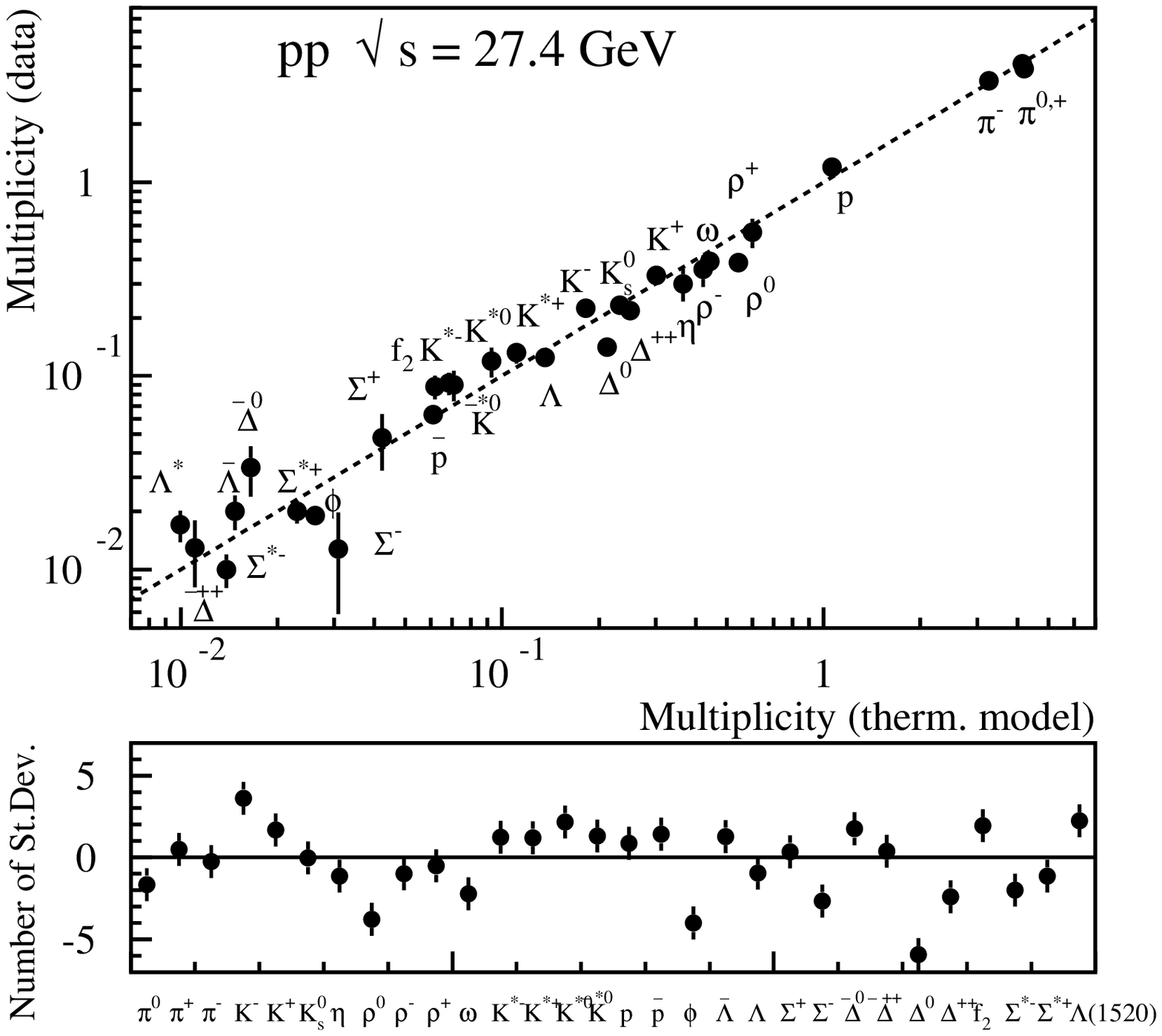,width=17cm}} 
\caption{}
\end{figure}

\newpage  
              
\begin{figure}[htbp]
\mbox{\epsfig{file=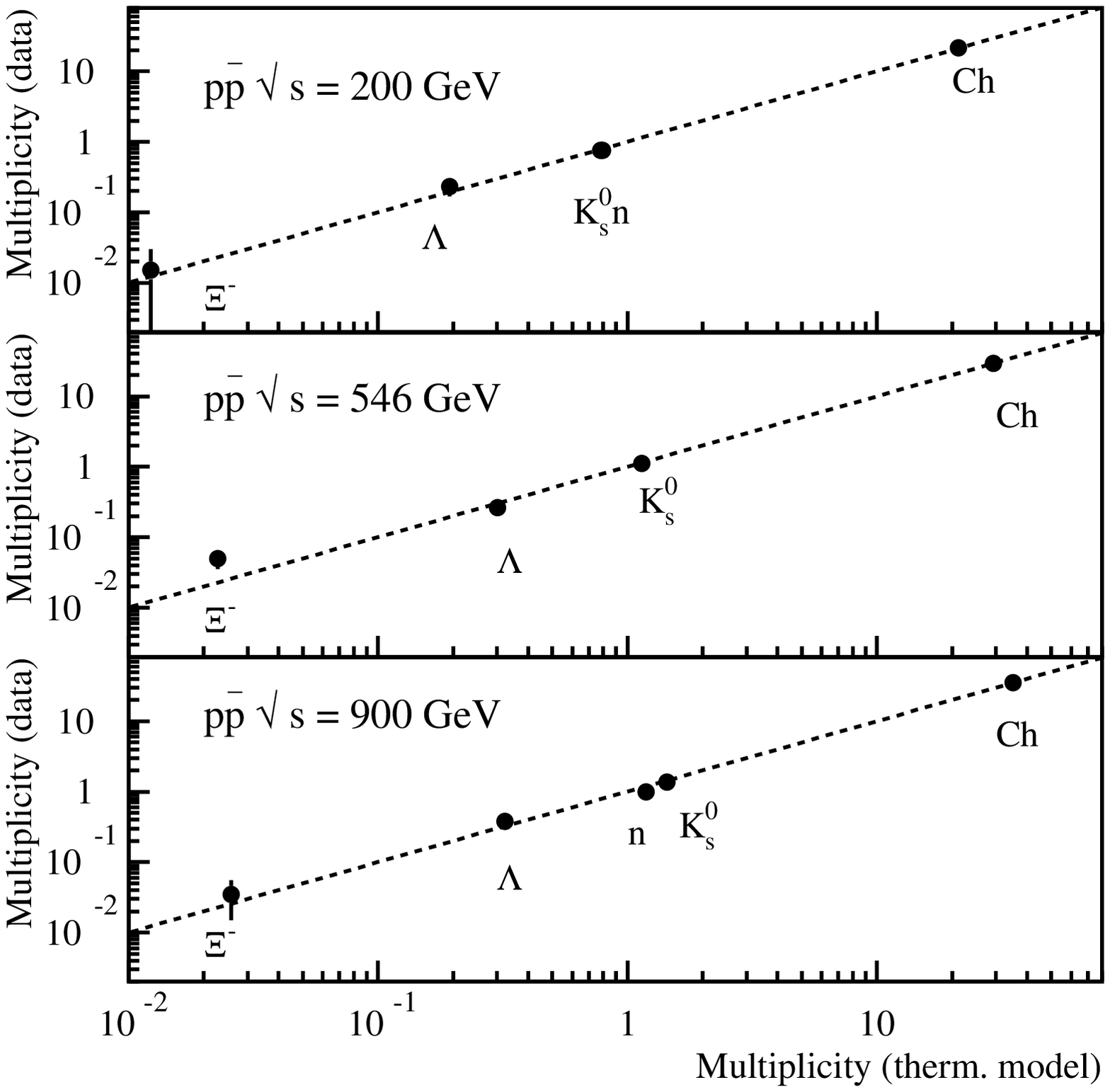,width=17cm}} 
\caption{}
\end{figure}

\newpage  
              
\begin{figure}[htbp]
\mbox{\epsfig{file=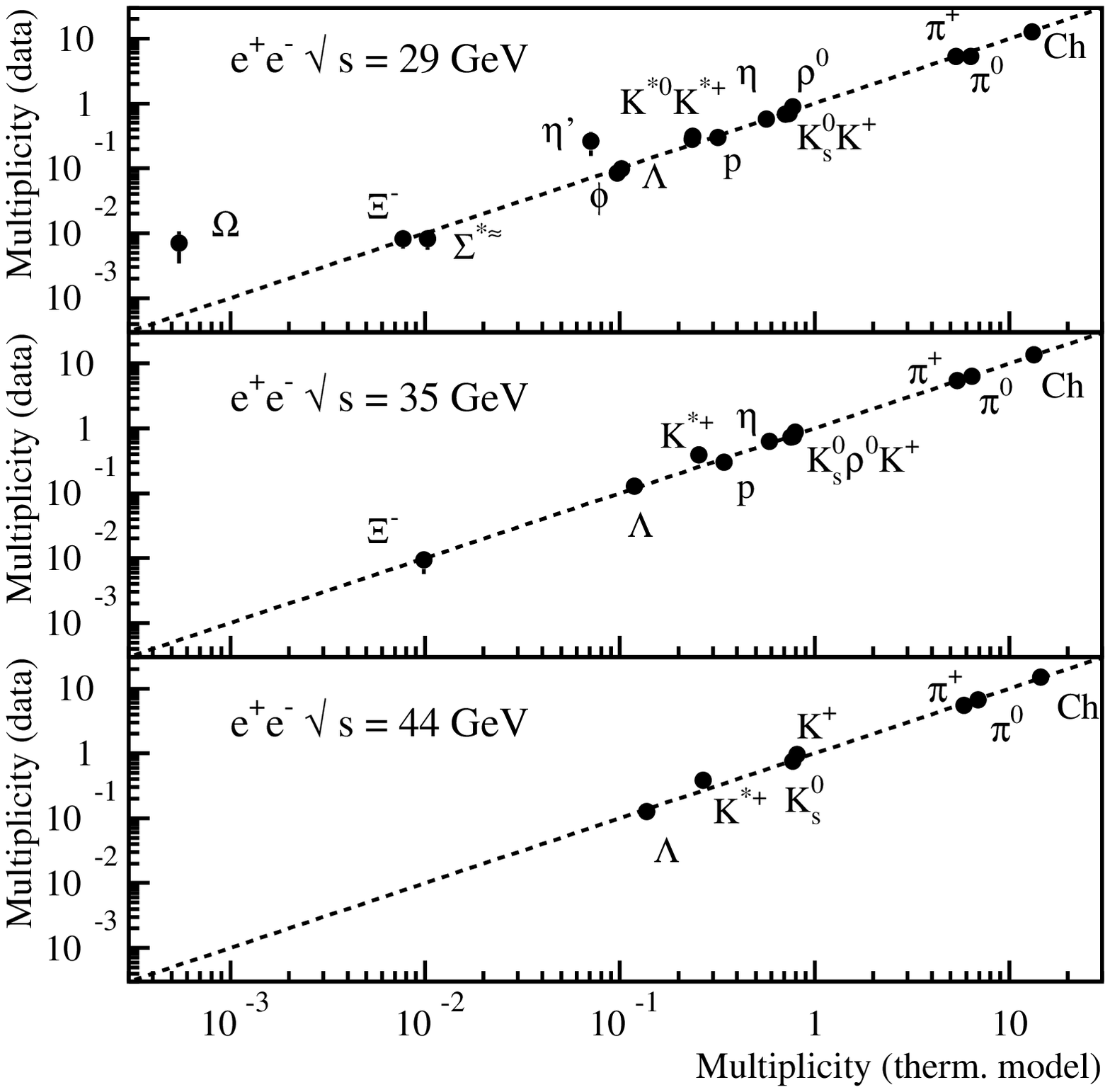,width=17cm}} 
\caption{}
\end{figure}

\newpage  
              
\begin{figure}[htbp]
\mbox{\epsfig{file=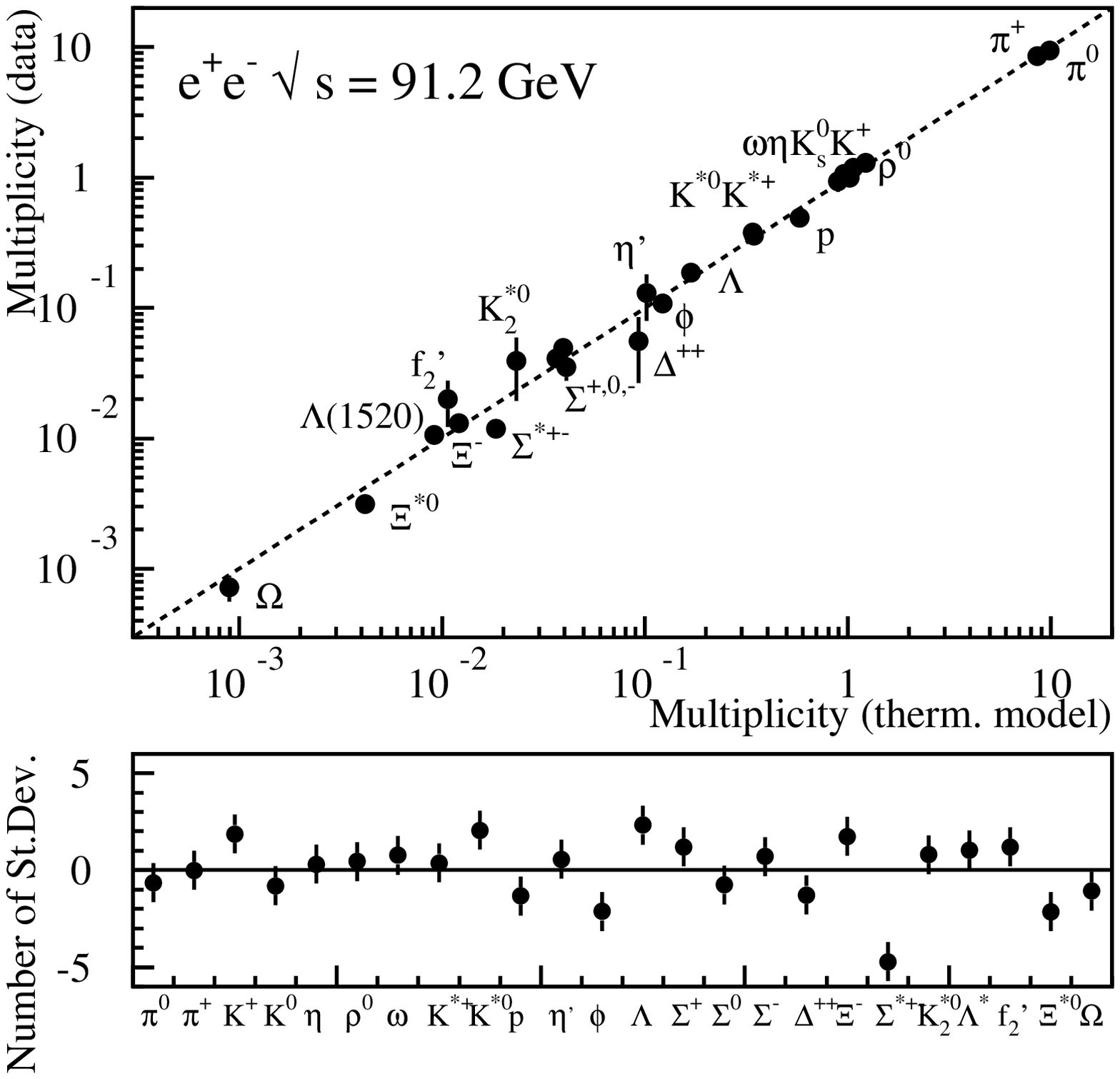,width=17cm}} 
\caption{}
\end{figure}

\newpage  
              
\begin{figure}[htbp]
\mbox{\epsfig{file=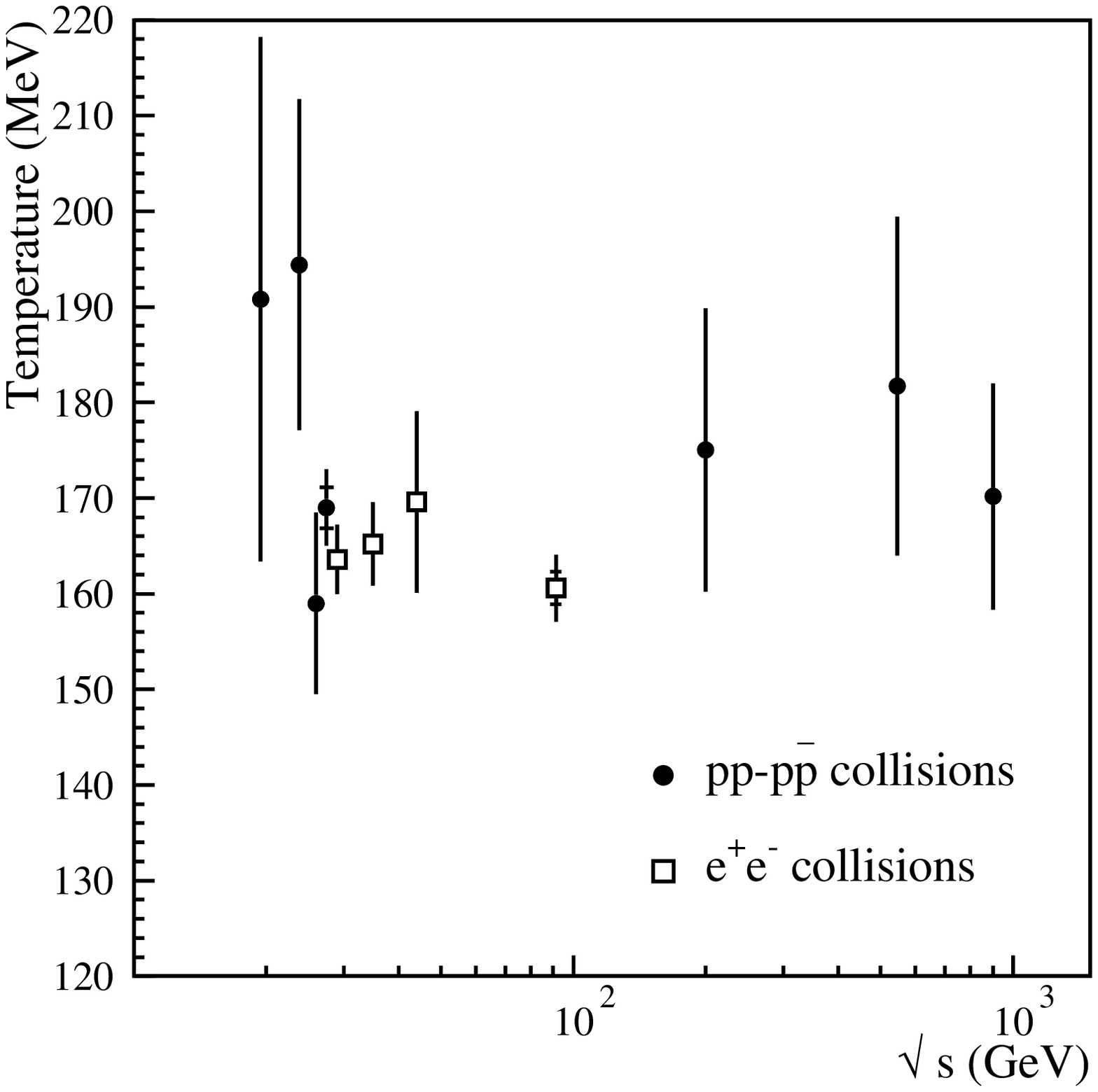,width=17cm}} 
\caption{}
\end{figure}

\newpage  
              
\begin{figure}[htbp]
\mbox{\epsfig{file=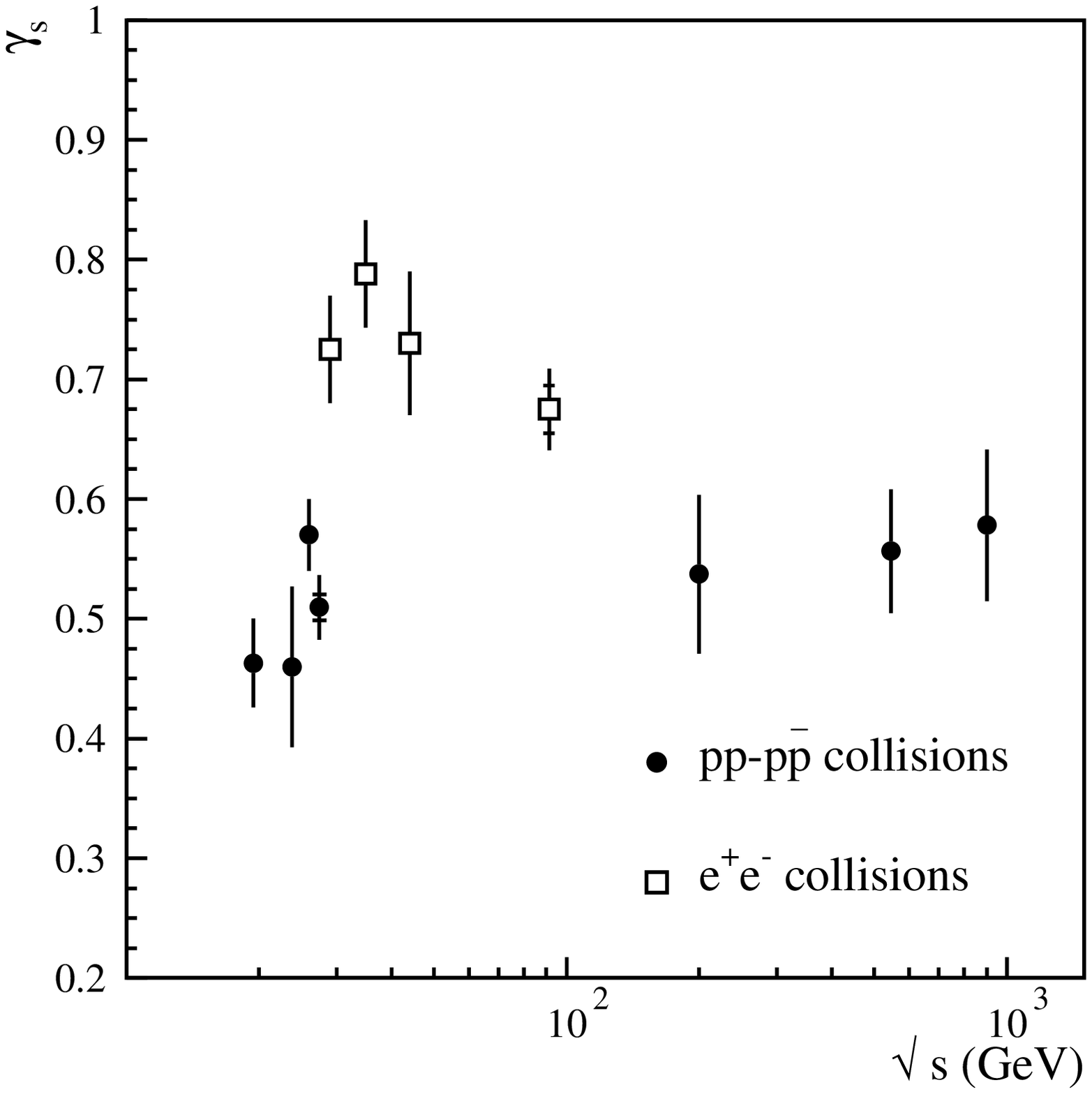,width=17cm}} 
\caption{}
\end{figure}

\newpage  
              
\begin{figure}[htbp]
\mbox{\epsfig{file=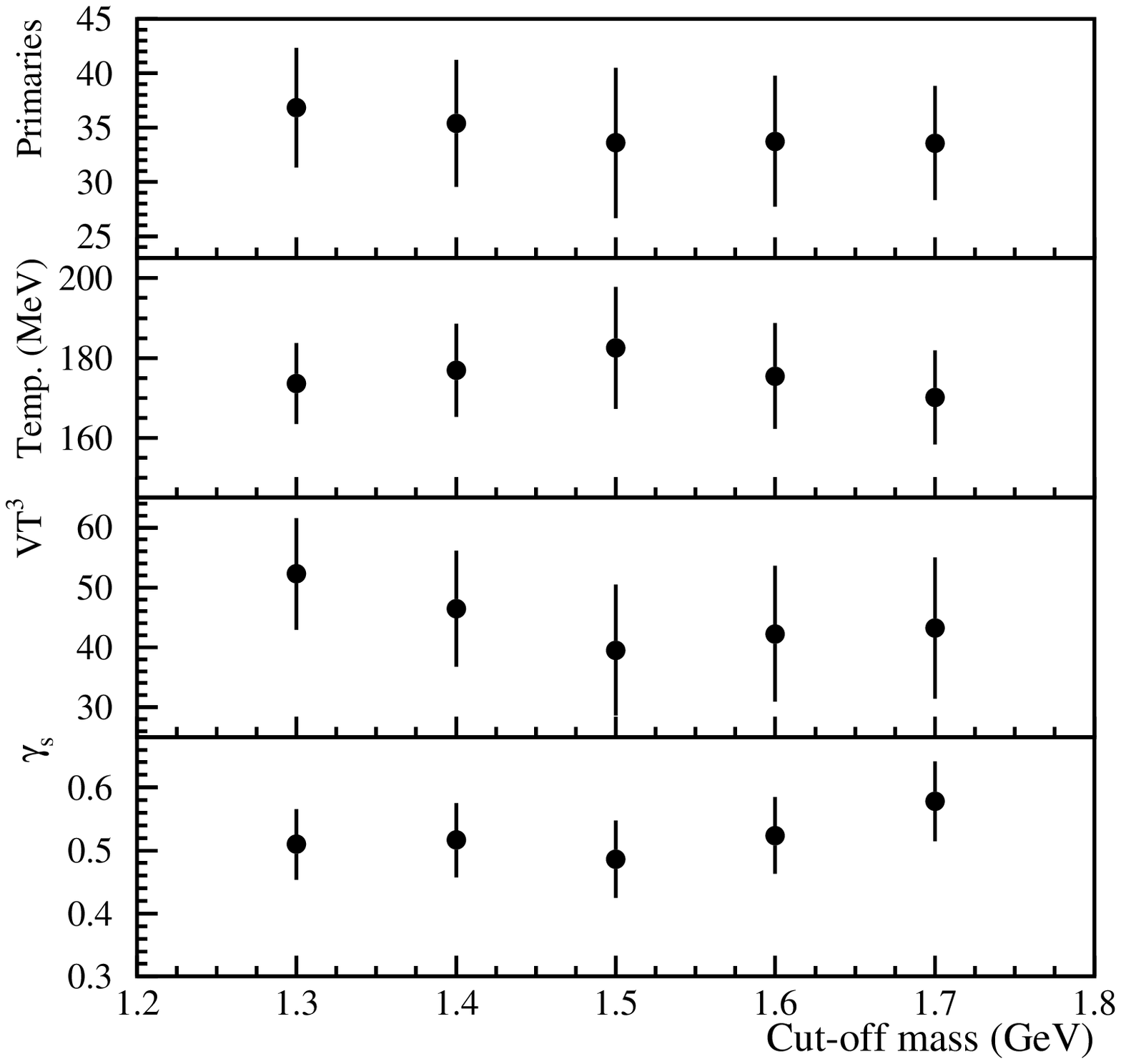,width=17cm}} 
\caption{}
\end{figure}

\end{document}